\documentclass[final,5p,times,twocolumn,authoryear]{elsarticle}

\makeatletter
\def\ps@pprintTitle{%
  \let\@oddhead\@empty
  \let\@evenhead\@empty
  \let\@oddfoot\@empty
  \let\@evenfoot\@oddfoot
}
\makeatother

\usepackage{amssymb}
\usepackage{lipsum}
\usepackage{color,soul}
\usepackage{flushend}
\usepackage{listings}
\usepackage{xcolor}
\usepackage{xcolor}
\usepackage{hyperref}
\usepackage{moresize}

\hypersetup{colorlinks=true,linkcolor=blue,urlcolor=blue}
\definecolor{codegreen}{rgb}{0,0.6,0}
\definecolor{codegray}{rgb}{0.5,0.5,0.5}
\definecolor{codepurple}{rgb}{0.58,0,0.82}
\definecolor{backcolour}{rgb}{0.95,0.95,0.92}

\lstdefinestyle{mystyle}{
  backgroundcolor=\color{backcolour}, commentstyle=\color{codegreen},
  keywordstyle=\color{magenta},
  numberstyle=\tiny\color{codegray},
  stringstyle=\color{codepurple},
  basicstyle=\ttfamily\scriptsize,
  breakatwhitespace=false,         
  breaklines=true,                 
  captionpos=b,                    
  keepspaces=true,                 
  numbers=left,                    
  numbersep=5pt,                  
  showspaces=false,                
  showstringspaces=false,
  showtabs=false,                  
  tabsize=2
}

\begin{document}

\begin{frontmatter}

\title{The Unreasonable Effectiveness of Monte Carlo Simulations in A/B Testing}

\author{Márton Trencséni (\texttt{mtrencseni@gmail.com})}

\begin{abstract}
This paper examines the use of Monte Carlo simulations to understand statistical concepts in A/B testing and Randomized Controlled Trials (RCTs). We discuss the applicability of simulations in understanding false positive rates and estimating statistical power, implementing variance reduction techniques and examining the effects of early stopping. By comparing frequentist and Bayesian approaches, we illustrate how simulations can clarify the relationship between p-values and posterior probabilities, and the validity of such approximations. The study also references how Monte Carlo simulations can be used to understand network effects in RCTs on social networks. Our findings show that Monte Carlo simulations are an effective tool for experimenters to deepen their understanding and ensure their results are statistically valid and practically meaningful.

\end{abstract}

\end{frontmatter}

\section{Introduction}

In his influential essay, \textit{The Unreasonable Effectiveness of Mathematics in the Natural Sciences} (Wigner, 1960), the Hungarian physicist Eugene Wigner marveled at the uncanny ability of abstract mathematics to describe and predict the physical world with remarkable precision. Wigner pondered how mathematical concepts, developed independently of empirical observations, could so effectively model the complexities of nature. This profound connection between mathematics and the natural sciences has been a cornerstone in advancing our understanding of the universe.

Building upon this theme, Halevy, Norvig, and Pereira's \textit{The Unreasonable Effectiveness of Data} (Halevy, Norvig and Pereira, 2009) emphasizes how vast amounts of data available today empower us to solve problems that were previously inaccessible. The paper argues that with sufficient data, simple models can outperform more sophisticated ones, highlighting the pivotal role of data over complex algorithms.

In a similar vein, Monte Carlo simulations are an effective tool across various scientific and engineering disciplines. Originating in the nuclear weapons work at Los Alamos and named after the famed casino city due to their reliance on randomness, Monte Carlo methods harness random sampling to approximate solutions to problems. However, their utility extends beyond tackling analytically intractable challenges. Monte Carlo simulations also allow us to directly simulate systems with known solutions, providing intuitive insights into why those solutions are what they are. By experimenting computationally, we can observe how changes in parameters affect outcomes, thereby deepening our understanding of the underlying processes—the computational experimentalist's calculus.

By running computational experiments that rely on random sampling, Monte Carlo simulations enable us to explore complex systems interactively. They serve as a bridge between theoretical models and practical applications, allowing us to validate analytical solutions, investigate their sensitivities, and intuitively grasp their implications. This approach is especially valuable in an era where computational resources are abundant and hands-on experimentation can complement traditional analytical methods.

\section{Historical context}

The origins of Monte Carlo methods date back to the 1940s during the development of nuclear weapons in the Manhattan Project. The mathematician Stanislaw Ulam conceived the idea in 1946 while recovering from an illness and playing solitaire: he wondered whether the chance of winning could be calculated by simulating numerous random deals rather than attempting a daunting combinatorial analysis. He developed the idea with John von Neumann for neutron diffusion calculations, and Nicholas Metropolis and Ulam later published \textit{The Monte Carlo Method} (Metropolis and Ulam, 1949). The term "Monte Carlo" was coined by Metropolis, referencing the famed casino city in Monaco, alluding to the randomness and chance inherent in the methods.

Since then, Monte Carlo methods have been adopted across a multitude of disciplines:

\begin{itemize}
\itemsep0em
  \item Physics: Modeling particle interactions, quantum mechanics, and statistical physics phenomena.
  \item Engineering: Solving integral equations, optimization problems, and reliability analysis.
  \item Finance: Pricing complex derivatives, risk assessment, and portfolio optimization.
  \item Computer Graphics: Rendering images using ray tracing and global illumination techniques.
  \item Statistics: Performing Bayesian inference and simulating sampling distributions.
\end{itemize}

The advent of modern computers significantly boosted the practicality of Monte Carlo simulations, enabling the handling of computations involving millions or billions of random samples. Monte Carlo methods are particularly effective due to several key factors. Firstly, their simplicity of implementation often requires fewer lines of code compared to analytical or deterministic numerical methods. By focusing on random sampling and aggregation, these methods reduce the need for intricate mathematical derivations, making them accessible to a broader range of practitioners.

Secondly, Monte Carlo simulations provide insight through experimentation. By directly simulating the system, researchers gain an intuitive understanding of how different factors influence outcomes. This experimental approach allows for the tweaking of parameters, observation of effects, and development of hypotheses—much like conducting physical experiments in a virtual environment. This hands-on aspect enhances comprehension of complex systems that might be analytically intractable.

Thirdly, the parallelization and computational efficiency of Monte Carlo methods make them particularly powerful in the modern computational landscape. The independent nature of random sampling allows simulations to be easily distributed across multiple processors. Modern multi-core CPUs, GPUs, and distributed computing systems can perform numerous simulations simultaneously, drastically reducing computation time. This efficiency enables the tackling of large-scale problems that were previously impractical, opening new avenues for research and application.

By leveraging these advantages, Monte Carlo simulations have become an indispensable tool across various scientific and engineering disciplines. They not only facilitate the approximation of solutions where traditional methods are infeasible but also democratize problem-solving by providing accessible tools that require less specialized mathematical background. This allows practitioners to focus on experimentation and exploration, driving innovation and deeper understanding in their respective fields.

The Python program below illustrates the simplicity of Monte Carlo by estimating the mathematical constant $\pi$ in essentially one line of code:

\begin{minipage}{\linewidth}
\begin{lstlisting}[language=Python, caption=Python code to estimate $\pi$. See the article \href{https://bytepawn.com/estimating-famous-mathematical-constants-with-monte-carlo-simulations.html}{Estimating mathematical constants with Monte Carlo simulations} for more examples and code samples.]
import math
import random

N = 100_000_000
pi_est = 4 * sum([1 for _ in range(N) if random.random()**2 + random.random()**2 < 1]) / N

print('Actual:   %.3f' % math.pi)
print('Estimate: %.3f' % pi_est)

> Actual:   3.142
> Estimate: 3.142
\end{lstlisting}
\end{minipage}

By using randomness, Monte Carlo simulations offer a powerful means to approximate solutions where traditional methods are infeasible or inefficient. They democratize problem-solving by providing accessible tools that require less specialized mathematical background, allowing practitioners to focus on experimentation and exploration.


\section{Randomized Controlled Trials and A/B testing}

Randomized Controlled Trials (RCTs) have been the cornerstone of experimental research for decades, particularly in fields such as medicine, psychology, and the social sciences. The fundamental principle behind RCTs is the random assignment of subjects to either a control group or a treatment group. This randomization aims to eliminate selection bias and confounding variables, so that, apart from chance variation, which can be quantified and controlled, observed differences between groups can be attributed to the intervention itself. By controlling for extraneous factors and utilizing random assignment, RCTs establish causal relationships between interventions and outcomes with a high degree of confidence.

In the medical field, RCTs are crucial for determining the efficacy and safety of new drugs, treatments, and medical procedures. For example, when testing a new medication, patients are randomly assigned to receive either the new drug or a placebo. Researchers then compare health outcomes between the two groups to assess the drug's effectiveness. The rigorous methodology of RCTs helps ensure that the results are reliable and can be generalized to broader populations.

With the advent of the internet and the proliferation of digital platforms, the principles of RCTs found a new application in the form of A/B testing (Kohavi, Tang and Xu, 2020). In this context, instead of patients and medical treatments, we deal with users and variations of digital content or features. A/B testing allows businesses and organizations to compare two versions of a webpage, app feature, email campaign, or any user interface element to determine which performs better according to predefined metrics such as click-through rates, conversion rates, time spent on site, or revenue generated.

For instance, an e-commerce website might want to test two different layouts for a product page to see which one leads to more purchases. Users are randomly assigned to see either version A or version B, and their interactions are tracked and analyzed. By comparing the performance metrics between the two groups, the business can make data-driven decisions about which design to implement. A/B testing empowers companies to optimize user engagement, enhance customer experience, and ultimately maximize revenues and profits. It embodies the experimental mindset in a business context, allowing for iterative improvements based on empirical evidence rather than intuition or subjective preferences.

However, the simplicity of setting up A/B tests belies the complexity of interpreting their results correctly. While modern analytics tools and platforms have made it easy to run experiments, ensuring that the conclusions drawn are valid requires a solid understanding of statistical principles. Several factors can compromise the integrity of an A/B test.

Misunderstanding statistical significance and disregarding statistical power is a common issue. A statistically significant result does not always imply a practically significant or meaningful effect. Confusion between p-values and effect sizes can lead to overestimating the importance of findings. Additionally, improper test design can introduce bias and invalidate results, for example lack of proper randomization between variants. Peeking at data and early stopping are other pitfalls; frequently checking the results of an ongoing test and stopping it once significance is achieved can inflate the false positive rate, as discussed later in the context of early stopping.

Monte Carlo simulations become invaluable in addressing these challenges. By providing a computational approach to modeling and analyzing the statistical properties of A/B tests, Monte Carlo methods allow experimenters to visualize statistical concepts and assess test validity. Simulating thousands of experiments enables the observation of the distribution of outcomes under various scenarios, making abstract statistical concepts more tangible. This approach helps identify potential pitfalls in experimental design by modeling how different factors, such as sample size or variance affect the likelihood of Type I and Type II errors.

Monte Carlo simulations also facilitate the optimization of test parameters. Experimenters can experiment with different configurations to determine optimal sample sizes, significance levels, and effect sizes needed to achieve desired statistical power. Understanding variance and uncertainty is enhanced by exploring how randomness and variability impact the results, leading to more robust interpretations and conclusions.

Implementing Monte Carlo simulations is facilitated by modern programming languages like Python, which offer accessible libraries for statistical modeling and random number generation. An experimenter can translate their experimental framework and assumptions into a simple Python program, running simulations that mirror the real-world A/B testing environment. This hands-on approach fosters a deeper understanding of the statistical mechanisms at play.

Moreover, Monte Carlo simulations can illustrate the consequences of common mistakes. By simulating the effects of early stopping, we can see how the false positive rate increases, reinforcing the importance of adhering to proper experimental protocols. Similarly, we can model the impact of variance reduction techniques, such as stratification or using covariates, to quantify their benefits in improving test sensitivity.

The remainder of this article will show specific examples of using Monte Carlo simulations to understand critical aspects of A/B testing. Through these examples, we will illustrate how computational experimentation enhances our ability to conduct robust experiments, interpret outcomes correctly, and make data-driven decisions with confidence.

\section{False positive rate and Statistical power}

Understanding false positives and statistical power is crucial in A/B testing, yet these concepts can sometimes feel abstract. Monte Carlo simulations offer a powerful way to visualize and comprehend these statistical phenomena by simulating thousands of experiments and observing their outcomes.

A false positive, or Type I error, occurs when we incorrectly conclude that there is a significant difference between two groups when, in reality, there isn't one. In the context of A/B testing, this means detecting an effect that doesn't exist.

To illustrate this, let's simulate an A/B test where both groups A and B have the same base conversion rate of 10\%. We run $ S = 10\,000 $ simulations, each time with a sample size of $ N = 1\,000\,000 $ users per group. The significance level $ \alpha $ is set at 0.05. We use a one-sided test: the alternative hypothesis is that B converts better than A.

\begin{minipage}{\linewidth}
\begin{lstlisting}[language=Python, caption=Python code to simulate A/B tests.]
# parameters
conversion_rate = 0.1    # true conversion rate
sample_size = 1_000_000  # number of samples
num_simulations = 10_000 # number of simulations
alpha = 0.05             # significance level

def compute_p_value_one_tailed(conversions_A, conversions_B, N):
    pooled_cr = (conversions_A + conversions_B) / (2 * N)
    s_error = np.sqrt(2 * pooled_cr * (1 - pooled_cr) / N)
    z_score = (conversions_B - conversions_A) / (N * s_error)
    p_value = 1 - stats.norm.cdf(abs(z_score))
    return p_value

# function to perform one A/B test simulation
def simulate_ab_test(conversion_rate, N, lift=0):
    conversions_A = np.random.binomial(N, conversion_rate)
    conversions_B = np.random.binomial(N, conversion_rate * (1 + lift))
    if conversions_B > conversions_A:
        return compute_p_value_one_tailed(conversions_A, conversions_B, N)
    else:
        return 1.0
\end{lstlisting}
\end{minipage}

Let's look at the scenario of an A/A test, when there is no difference between the two variants, the lift is 0\%. Any statistically significant difference detected is a false positive.

\begin{minipage}{\linewidth}
\begin{lstlisting}[language=Python, caption=A/A test False Positive Rate (FPR) simulation.]
# A/A test, ie. A and B are the same
lift = 0.0
p_values = [simulate_ab_test(
    conversion_rate,
    sample_size,
    lift) for _ in range(num_simulations)]
fpr = np.mean(np.array(p_values) < alpha)
print(f"False positive rate: {fpr:.3f}")

> False positive rate: 0.051
\end{lstlisting}
\end{minipage}

Out of $ 10\,000 $ simulations, approximately 5\% resulted in p-values less than 0.05. This aligns with our significance level $ \alpha = 0.05 $, confirming that the probability of a Type I Error is controlled by $ \alpha $. Even when there is no true difference, we expect to see some false positives purely due to chance. Next, we simulate an A/B test where the treatment group has a 1\% relative improvement in the conversion rate. In A/B testing, statistical power is the ratio of cases when we detect an effect, if the effect is there. Unlike the false positive rate, which we control explicitly with the significance level $ \alpha $, statistical power is a function of the sample size, the effect size and $ \alpha $. We can easily use the same Monte Carlo framework to measure statistical power in this experiment setup:

\begin{minipage}{\linewidth}
\begin{lstlisting}[language=Python, caption=A/B test statistical power computation.]
# A/B test, B is 1% better than A
lift = 0.01
p_values = [simulate_ab_test(
    conversion_rate,
    sample_size,
    lift) for _ in range(num_simulations)]
sp = np.mean(np.array(p_values) < alpha)
print(f"Statistical power: {sp:.3f}")

> Statistical power: 0.762
\end{lstlisting}
\end{minipage}

The statistical power is about 76\%, indicating about a 3 in 4 chance of detecting a true effect of 1\% lift. The risk of not detecting the effect when it's in fact there (false negative, Type II Error) is about 1 in 4 in this scenario. With a two-sided test, the statistical power in the same setup would be about 65\%.

These simple simulations highlight the basic effectiveness of Monte Carlo methods. By simulating thousands of experiments, we observe how random chance can lead to false positives and how true effects are detected; we confirm the observed Type I Error rate matches the specified significance level. Statistical power was not explicitly specified in the parameters, it is a function or result of the chosen values and estimated with Monte Carlo simulations. The experimenter can then vary the sample size to see how to achieve a potentially desired higher statistical power at a given assumed lift.

These short snippets illustrate how Monte Carlo simulations make abstract statistical concepts tangible, helping us grasp the real-world implications of our A/B testing strategies.

\section{Variance Reduction Techniques}

In A/B testing, accurately detecting differences between two groups hinges on minimizing variance in the measurements. When we compare the means of a metric—such as user spend or conversion rates—across control and treatment groups, sample sizes of a few thousand units per variant usually suffice for the Central Limit Theorem to apply, although heavy-tailed metrics such as spend can require more, meaning that the sampling distribution of the mean approximates a normal distribution. Consequently, the difference between the two group means (often referred to as the lift) also follows a normal distribution, with its variance being the sum of the variances of the two group means. Reducing this variance is crucial for increasing the precision and reliability of our experimental results. See the article \href{https://bytepawn.com/ab-testing-and-the-central-limit-theorem.html}{A/B testing and the Central Limit Theorem} for a more in-depth exploration of the CLT with Monte Carlo methods.

\begin{figure}[h]
  \centering 
  \includegraphics[width=0.35\textwidth]{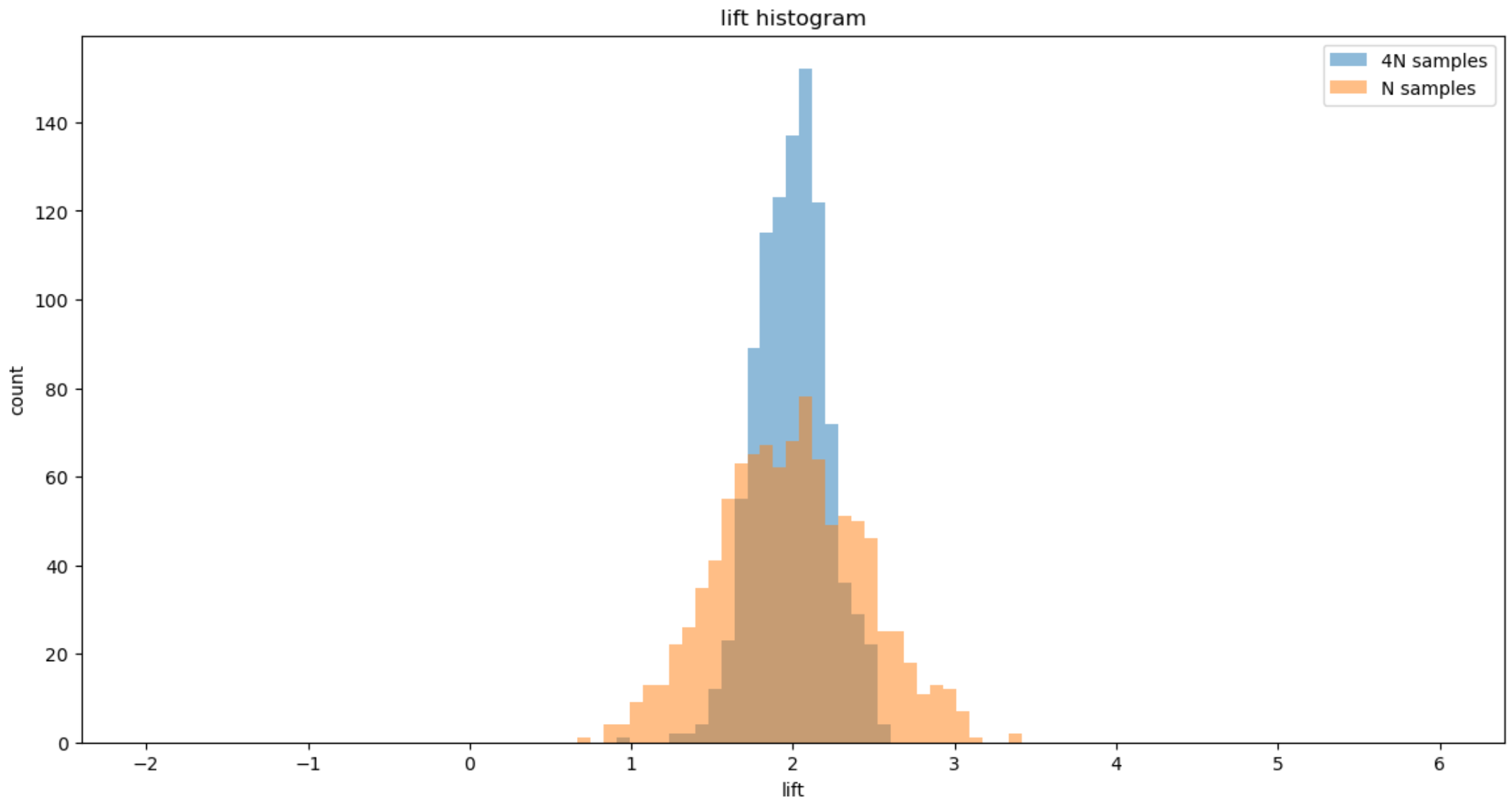}  
  \caption{Monte Carlo simulation illustrating that increasing the sample size reduces the variance of the lift measurement.}
  \label{fig:samplesize}
\end{figure}

The most straightforward way to reduce variance is by increasing the sample size. According to statistical principles, the variance of the sample mean is inversely proportional to the sample size. Specifically, quadrupling the number of observations halves the standard error, as shown in Figure~\ref*{fig:samplesize}. However, collecting more data may not be feasible due to time constraints, resource limitations, or the finite availability of participants. Additionally, extending the duration of an experiment doesn't always yield a proportional increase in unique users, as returning visitors may inflate sample counts without providing new information.

\begin{figure}[!b]
  \centering 
  \includegraphics[width=0.35\textwidth]{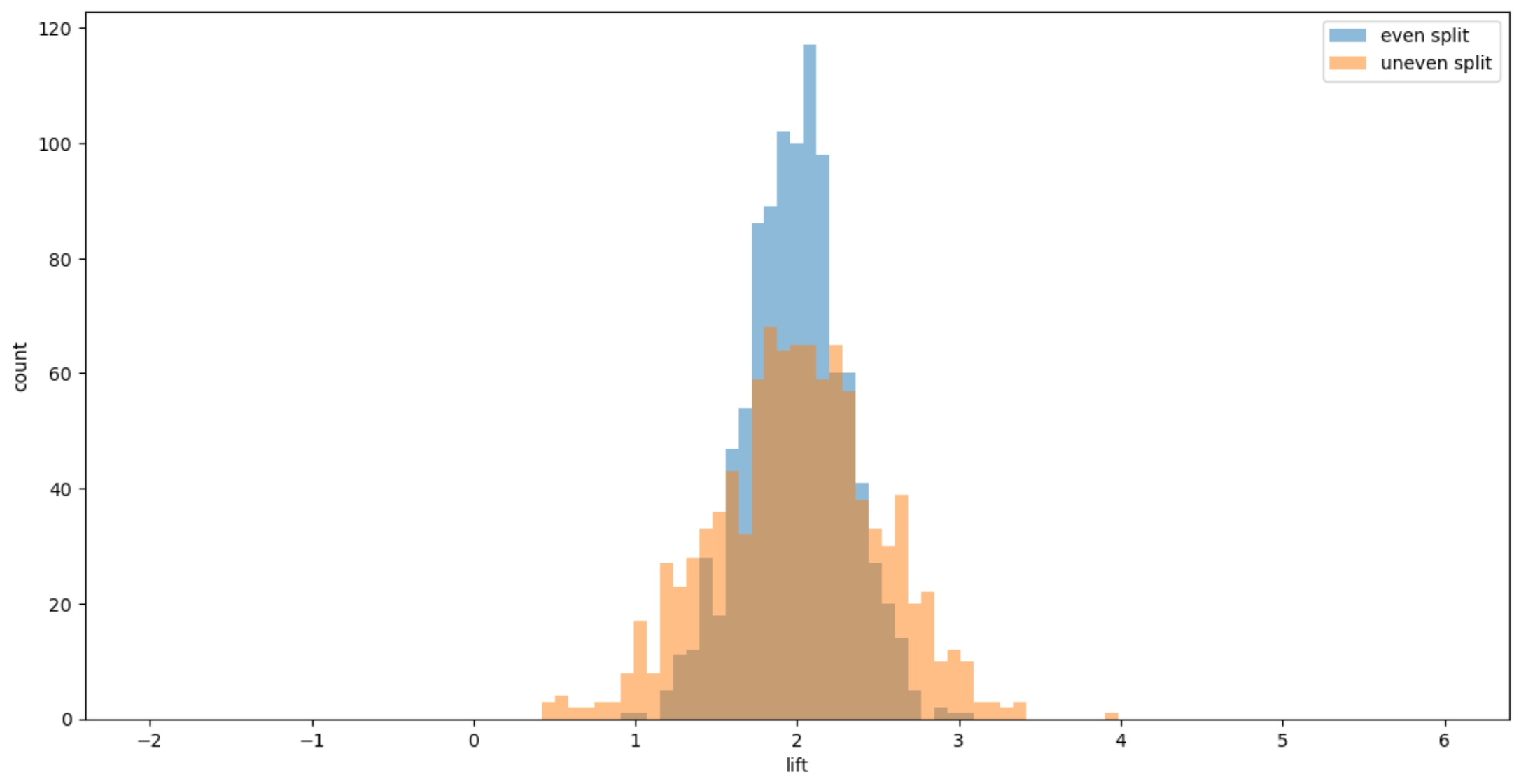}  
  \caption{Monte Carlo simulation illustrating that moving towards a more even split reduces the variance of the lift measurement.}
  \label{fig:split}
\end{figure}

Allocating participants equally between the control and treatment groups minimizes variance in the estimated effect size. When the total sample size is fixed and the two groups have equal variance, a 50\%-50\% split ensures the lowest possible variance for the difference between group means. Deviations from an even split increase variance because the variance of each group's mean is inversely proportional to its sample size. An uneven allocation, such as 90\%-10\%, results in a higher combined variance compared to an even split, as illustrated in Figure~\ref*{fig:split}. Therefore, when feasible, it's advisable to distribute participants equally to enhance the sensitivity of the experiment. With unequal variances, the optimal split allocates samples in proportion to each group's standard deviation. However, in practice an even split is usually not chosen, as risk control considerations lead to significantly fewer users in the risky treatment group. Risk control considerations have to be consciously weighed and balanced with the possibility of lower variance by the experiment designers.

Another approach to variance reduction involves refining the metric itself to make it less variable. This can be achieved by winsorizing or trimming outliers, where extreme values that disproportionately inflate variance and obscure true effects are capped or removed beyond a certain threshold, thereby focusing the analysis on the central tendency of the data; for instance, in measuring spend per user, extraordinarily high purchases—possibly due to anomalous behavior or errors—can be capped to reduce their impact. Using medians instead of means is another method, as the median is less sensitive to extreme values and provides a more robust measure when the data distribution is skewed.\footnote{The median does not always have a lower variance, and may necessitate different statistical tests or non-parametric methods. Trimming, winsorizing and switching to medians change the quantity being estimated, so they should be chosen before the experiment and interpreted accordingly.} Additionally, focusing on relevant subpopulations by segmenting users into more homogeneous groups based on characteristics like engagement level or purchase history can reduce within-group variance, yielding more precise estimates when these subgroups are analyzed separately. Lastly, splitting metrics by decomposing a complex metric into its components can reduce variance; for example, overall spend per user can be divided into the conversion rate—the proportion of users who make a purchase—and the average spend among purchasers, with each component typically exhibiting lower variance than the combined metric and offering clearer insights when analyzed individually.

Stratified sampling involves dividing the overall sample into distinct subgroups or strata based on specific characteristics, such as age, gender, or previous behavior, and ensuring that these strata are represented in the same proportions in both control and treatment groups. This method reduces variance by removing the variability between these subgroups that would otherwise come from imbalances in how they randomly fall into control and treatment. By ensuring that each subgroup contributes proportionally to both groups, we eliminate potential sources of variance unrelated to the treatment effect. Stratification is particularly effective when the outcome metric varies significantly across these subgroups.

Incorporating covariates—additional variables that are correlated with the outcome metric—can significantly reduce variance. CUPED (Controlled Experiment Using Pre-Experiment Data) (Deng, Xu, Kohavi and Walker, 2013) is a technique that adjusts for variance using covariates measured before the experiment begins. If historical data on the outcome metric is available, and there is a strong correlation between past and current behavior, this information can be used to adjust the post-treatment measurements.

\begin{figure}[h]
  \centering 
  \includegraphics[width=0.35\textwidth]{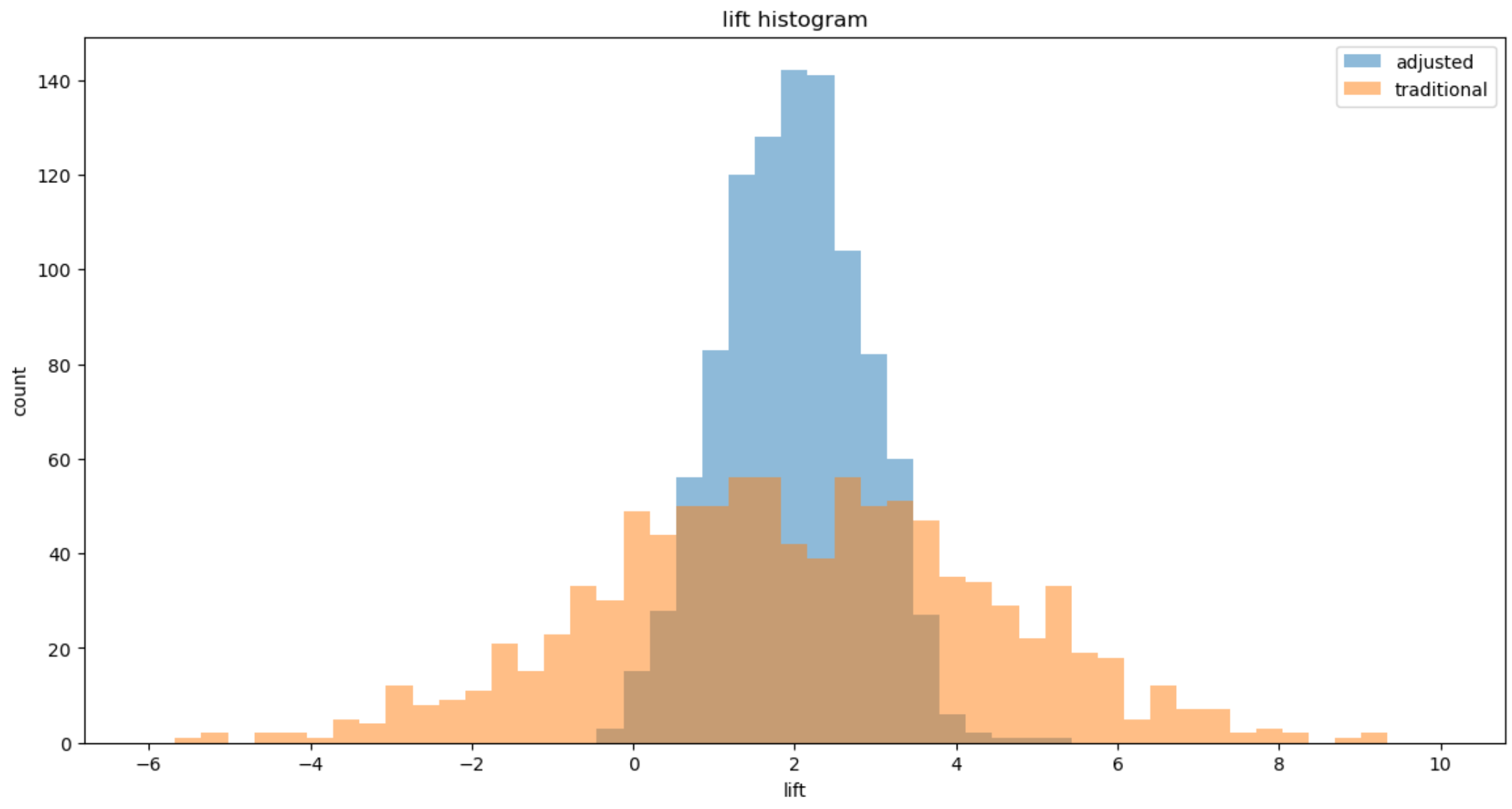}  
  \caption{Monte Carlo simulation illustrating that using CUPED to take into account historic data reduces the variance of the lift measurement.}
  \label{fig:cuped}
\end{figure}

CUPED works by computing an adjusted metric that accounts for the covariance between the pre-experiment and post-experiment data. This adjustment effectively removes the portion of variance explained by the covariate, leading to a more precise estimate of the treatment effect, as illustrated in Figure~\ref*{fig:cuped}. For example, if users who spent more before the experiment tend to spend more during the experiment, using their prior spend as a covariate can control for this variability.

See the article \href{https://bytepawn.com/five-ways-to-reduce-variance-in-ab-testing.html}{Five ways to reduce variance in A/B testing} for more information and code on variance reduction techniques, explored using Monte Carlo simulations.

\section{Early stopping}

Early stopping in A/B testing is a critical consideration that can significantly impact the validity of experimental results. In a standard experimental protocol, we begin by defining the evaluation metric—selecting a key performance indicator such as conversion rate, time spent, or daily active users. Next, we determine the sample size $ N $ required to achieve sufficient statistical power, choose the appropriate statistical test based on the data characteristics (such as a t-test or chi-squared test), and set the significance level $ \alpha $, which establishes the acceptable false positive rate, commonly at 0.05. We then collect data by running the experiment until the predetermined sample size is reached and analyze the results by performing the statistical test and interpreting the p-value to reject or fail to reject the null hypothesis.

However, in practice, experimenters often face the temptation of early stopping—the act of peeking at the data before the experiment reaches the planned sample size and making decisions based on interim results. While this may seem like a way to expedite decision-making, early stopping can significantly inflate the false positive rate, leading to incorrect conclusions.

Early stopping disrupts the statistical integrity of the experiment. Each time we perform an interim analysis and test for significance, there's a chance of observing a false positive purely by random chance. If we repeatedly check the data and stop the experiment once we achieve statistical significance, we inadvertently increase the overall probability of committing a Type I Error (false positive). For instance, suppose we plan to collect $ N = 3\,000 $ samples but decide to peek at the results every $ 1\,000 $ samples. At each interim analysis, we perform a statistical test at the significance level $ \alpha = 0.05 $. The cumulative effect of conducting multiple tests increases the overall false positive rate beyond the nominal 5\%. In fact, simulations show that with three such analyses, the false positive rate roughly doubles, to about 11\%.

\begin{figure}[h]
  \centering 
  \includegraphics[width=0.35\textwidth]{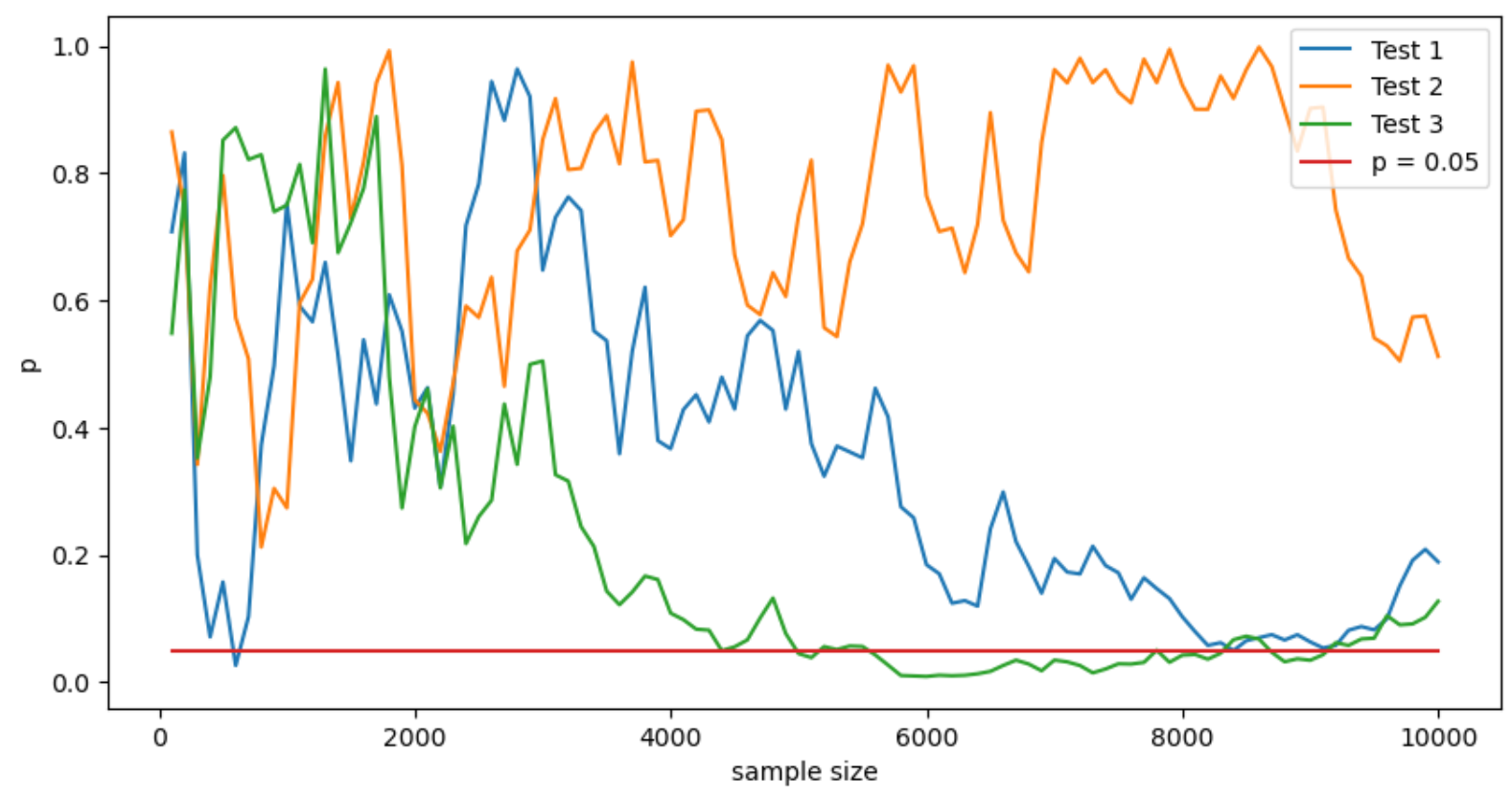}  
  \caption{P-values for three A/A test simulation runs, evaluated every 100 samples, 100 times for a total of $ N=10\,000$ samples. 2 out of 3 Monte Carlo runs cross the critical $\alpha$ level before the experiment concludes, even though all 3 end up without hitting statistical significance.}
  \label{fig:earlystopping}
\end{figure}

The intuition behind this increase lies in the nature of p-values and random variability. Under the null hypothesis (no real difference between variants A and B), p-values are uniformly distributed between 0 and 1. This means that at any given time, there's always a probability $ \alpha $ of observing a p-value less than the significance threshold $ \alpha $ purely by chance. When we perform multiple interim analyses, each one provides an additional opportunity to cross the significance threshold due to random fluctuations in the data. The more we peek, the higher the chance that we'll observe a statistically significant result when there is no true effect. This is shown in Figure~\ref*{fig:earlystopping}, illustrating the p-values for three A/A test simulation runs, evaluated every 100 samples, 100 times for a total of $ N=10\,000$ samples. These simulations demonstrate that the overall false positive rate increases substantially compared to a single analysis at the end of the experiment, highlighting the risks associated with early stopping.

To mitigate the increased false positive rate while allowing for interim analyses, statisticians have developed methods to control the overall Type I error rate. One such approach is group sequential testing, which distributes the total allowable Type I error ($ \alpha $) across multiple analyses; $ \alpha $ spending functions (Lan and DeMets, 1983) generalize this to arbitrary analysis schedules. For example, Pocock's boundary (Pocock, 1977) uses the same, more stringent significance level at every analysis, including the final one. If we plan three equally spaced analyses and want to maintain an overall two-sided $ \alpha = 0.05 $, we set the significance level at each analysis to $ \alpha' = 0.0221 $. This adjustment ensures that the cumulative probability of a Type I error across all analyses remains at the desired level.

Another approach is the Haybittle–Peto boundary (Haybittle, 1971; Peto et al., 1976), which sets a very stringent significance level (e.g., $ \alpha = 0.001 $) for interim analyses and retains the standard $ \alpha = 0.05 $ for the final analysis. This method is simple and keeps the overall false positive rate close to the nominal level (slightly above it, about 5.05\% for three equally spaced analyses) without complex calculations. Monte Carlo simulations can validate these adjusted thresholds by demonstrating that the overall false positive rate remains at or near the intended level when these methods are applied.


Early stopping in A/B testing poses a significant risk to the validity of experimental results by inflating the false positive rate. While the allure of quick wins is strong, especially in fast-paced business environments, maintaining statistical rigor is essential for making reliable data-driven decisions. Understanding the underlying concepts of A/B testing and leveraging tools like Monte Carlo simulations enhances the effectiveness of experiments and contributes to a statistically sound long-term decision-making process. See the article \href{https://bytepawn.com/early-stopping-in-ab-testing.html}{Early stopping in A/B testing} for a more in-depth explanation with code samples and figures.

\section{Frequentist and Bayesian A/B testing}

In the practice of A/B testing, two predominant statistical frameworks guide our analysis: frequentist statistics and Bayesian inference. Both aim to determine whether a new treatment variant (B) performs differently from a control (A), but they approach the problem from different philosophical standpoints and methodologies.

The frequentist approach (which we have used up until now) revolves around the formulation of hypotheses. We set up a null hypothesis $ H_0 $, which typically states that there is no difference between the conversion rates of A and B—that any observed difference is due to random chance. The alternative hypothesis $ H_1 $ (sometimes called the action hypothesis) posits that there is a difference, with B performing better than A. Since $ H_1 $ is directional, the p-value here is one-sided. After running the experiment, we calculate a p-value $ p_f $ ($f$ for frequentist), which represents the probability of observing data at least as extreme as the actual results, assuming the null hypothesis is true. If this p-value is lower than a predetermined significance level (commonly $ \alpha = 0.05 $ or $ 0.01 $), we reject the null hypothesis in favor of the alternative, concluding that the treatment has a statistically significant effect.

On the other hand, the Bayesian approach flips this perspective. Instead of focusing on the probability of the data given the null hypothesis, it computes the probability that the alternative hypothesis is true given the observed data. This involves using Bayes' theorem to update our prior beliefs about the parameters of interest based on the experimental outcomes, resulting in a posterior probability $ p_b $ ($b$ for Bayesian). In Bayesian A/B testing, we model the conversion probabilities of A and B using probability distributions, often choosing the Beta distribution due to its suitability for modeling probabilities between 0 and 1.

The Beta distribution is parameterized by two parameters, $ a $ and $ b $, which with a flat prior correspond to one plus the number of successes (conversions) and one plus the number of failures (non-conversions), respectively. Given the observed data, say $ C $ conversions out of $ N $ trials, we set $ a = C + 1 $ and $ b = N - C + 1 $ for each variant. This allows us to express our uncertainty about the true conversion rate of each variant as a probability distribution. To determine the probability that B is better than A, we compute $ P(\mu_B > \mu_A) $, where $ \mu_A $ and $ \mu_B $ are the true conversion rates of A and B. This involves integrating the joint probability distribution over the region where $ \mu_B > \mu_A $. While sometimes a closed-form solution exists, we often use Monte Carlo integration by sampling from the Beta distributions of both A and B and estimating the proportion of times $ \mu_B > \mu_A $.

\begin{figure}[h]
  \centering 
  \includegraphics[width=0.35\textwidth]{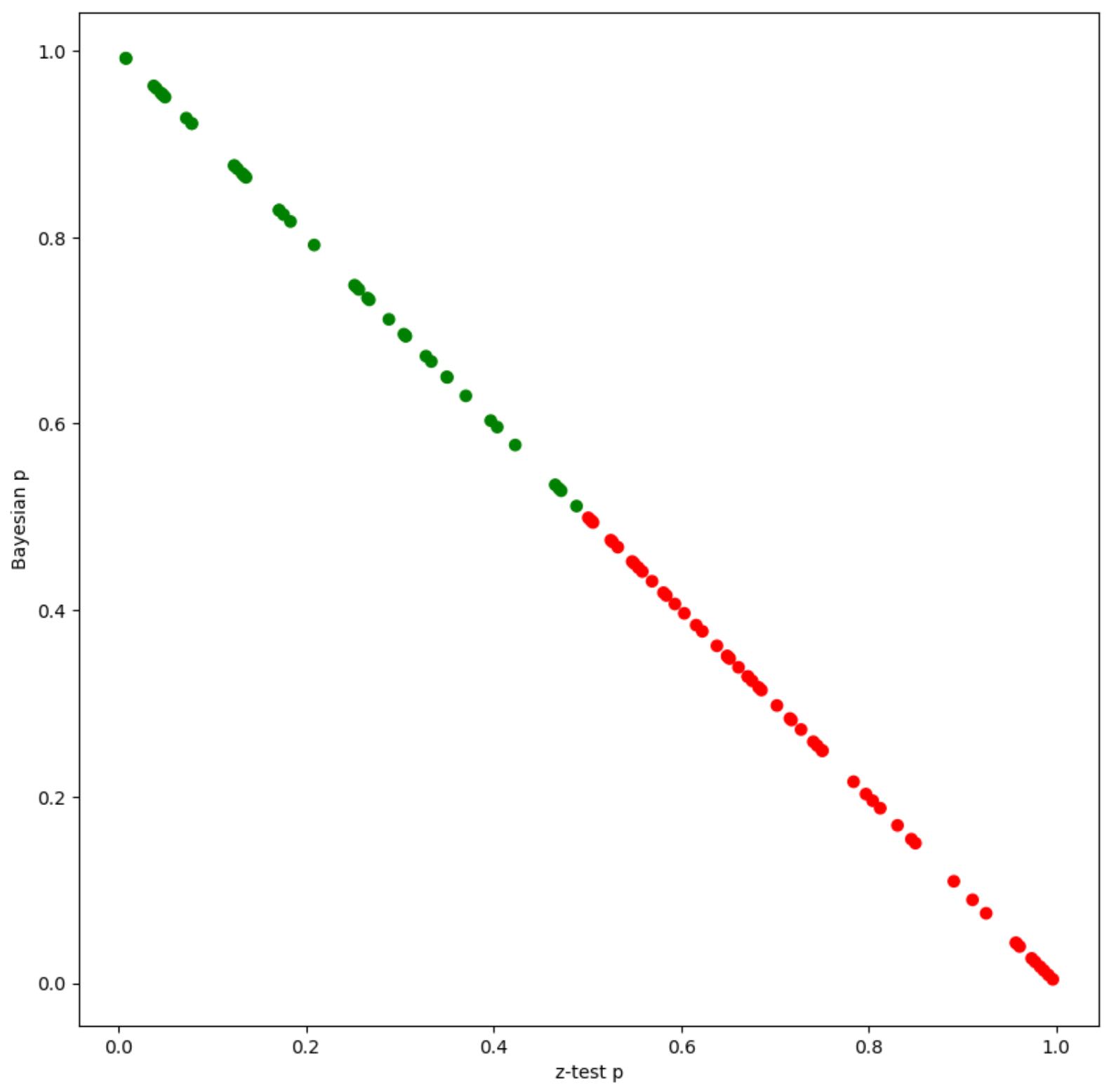}  
  \caption{Monte Carlo simulations show that in many scenarios, the frequentist one-sided p-value $p_f$ and the Bayesian probability $p_b$ are complementary and the approximation $ p_f + p_b \approx 1 $ holds.}
  \label{fig:pfpb}
\end{figure}

An intriguing observation emerges when comparing the frequentist p-value $ p_f $ and the Bayesian posterior probability $ p_b $. Under certain conditions—specifically, when using flat priors and with sufficiently large sample sizes—the relationship $ p_f + p_b \approx 1 $ holds, as illustrated in Figure~\ref*{fig:pfpb} using Monte Carlo methods. This relationship requires the one-sided p-value. With the more common two-sided p-value, $ p_f/2 + p_b \approx 1 $ when B's observed conversion rate is higher than A's, and $ p_f/2 \approx p_b $ when it is lower. This suggests that, in these scenarios, the frequentist and Bayesian methods provide complementary perspectives on the same data. However, this is an approximation and does not hold exactly, especially with small sample sizes or strong prior beliefs.

To understand these approximate equalities, consider that in frequentist inference, particularly when using the z-test, we rely on the Central Limit Theorem to approximate the sampling distribution of the conversion rates as normal distributions. The difference in conversion rates between A and B is also normally distributed; under the null hypothesis it is centered on zero, with a variance that depends on the individual variances of A and B. We then calculate the probability of observing a difference at least as extreme as the one observed, assuming the null hypothesis is true. In Bayesian modeling, when we use Beta distributions with weak or non-informative priors, and when sample sizes are large enough and conversion rates not too close to 0 and 1, the Beta distributions approximate normal distributions centered on the observed conversion rates, with nearly the same variances as the frequentist normal approximations, so the posterior probability that B is better than A approximately equals one minus the one-sided p-value. This explains why $ p_f + p_b \approx 1 $ holds under these conditions. However, the choice of prior in Bayesian analysis can significantly impact the results. If we use a strong prior, such as $ Beta(a_0, b_0) $ with large $ a_0 $ and $ b_0 $, our prior beliefs may dominate the posterior distribution, requiring substantial new data to shift our confidence. This means that with strong priors, the approximation $ p_f + p_b \approx 1 $ may not hold, and the Bayesian probability $ p_b $ may remain relatively unchanged despite new data.

Monte Carlo simulations play a crucial role in bridging these concepts. By simulating a large number of experiments, we can visualize how the Bayesian posterior probability $ p_b $ and the frequentist p-value $ p_f $ behave under different conditions. For instance, with small sample sizes or extreme conversion rates (close to 0 or 1), the Beta distributions differ significantly from the normal approximations, and the relationship between $ p_f $ and $ p_b $ becomes less predictable. See the article \href{https://bytepawn.com/bayesian-ab-conversion-tests.html}{Bayesian A/B conversion tests} for an in-depth explanation of the concepts above using Monte Carlo simulations.

\section{Social Networks}

In traditional A/B testing, a critical assumption is that individual units—users, customers, or participants—are independent of one another. This Independence Assumption (IA) implies that the outcome of one unit does not influence the outcome of another. However, in the context of social networks, this assumption is often false. Users are interconnected, influencing each other's behaviors through interactions, content sharing, and social influence. This interconnectedness introduces network effects that can significantly impact the results of A/B tests.

By modeling user behavior over simulated social networks, we can explore and quantify various phenomena that affect A/B testing in these interconnected environments. These simulations help us understand how traditional statistical methods may misrepresent true effects when the IA does not hold, allowing us to adjust our approaches accordingly.

Consider a scenario where we aim to test a new feature designed to increase user content production on a social platform. Users are connected in a network, and their activity influences the activity of their friends. If a user in the treatment group starts producing more content due to the new feature, their friends in the control group might also produce more content upon seeing this increased activity. This phenomenon is known as the spillover effect.

\begin{figure}[h]
  \centering 
  \includegraphics[width=0.35\textwidth]{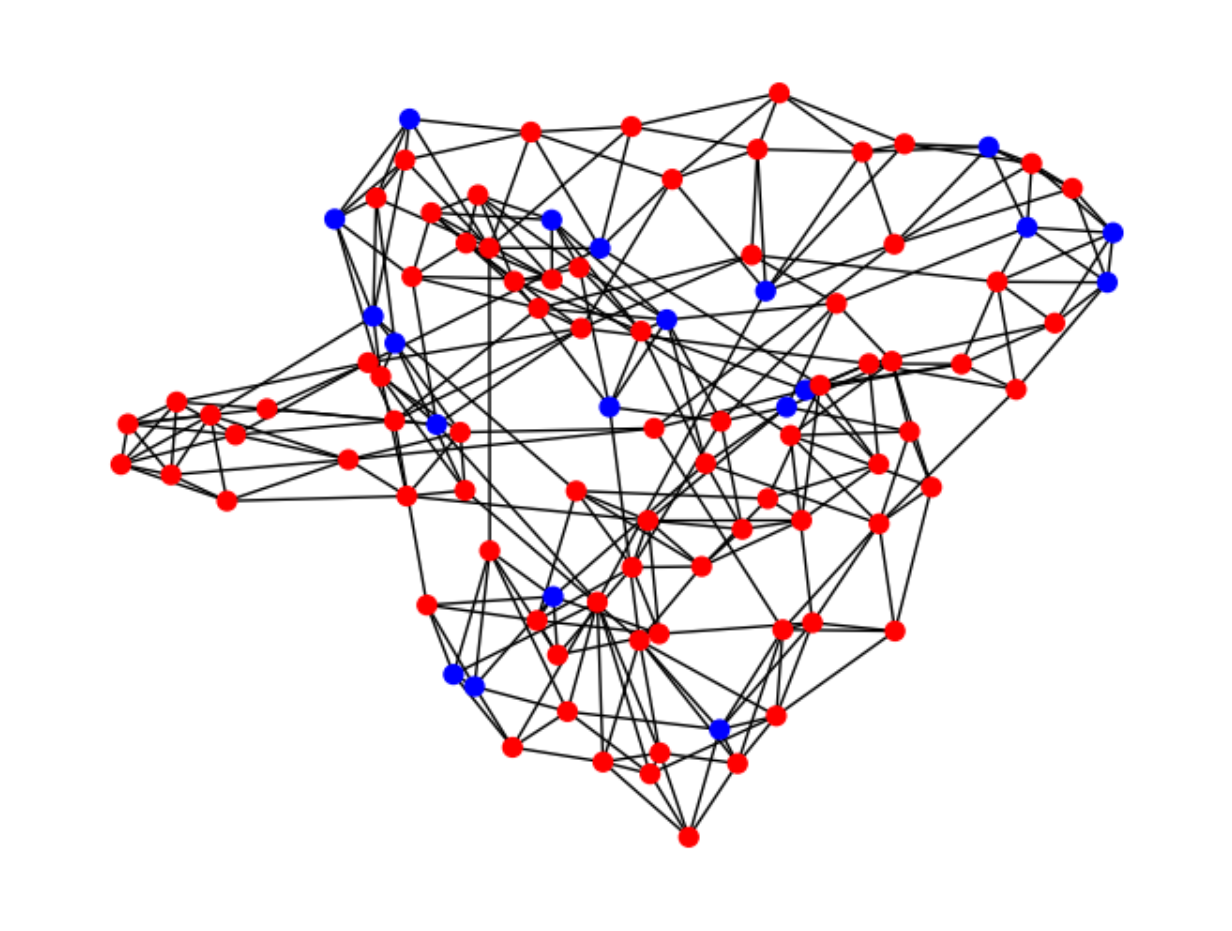}  
  \caption{Watts-Strogatz graph with $(n=100, k=6, p=0.1)$, with $N=20$ random treatment nodes colored blue.}
  \label{fig:network}
\end{figure}

Monte Carlo simulations allow us to model such interactions by creating synthetic social networks—for example using models of networks like Watts-Strogatz graphs (Watts and Strogatz, 1998), as in Figure~\ref*{fig:network}—and simulating user behavior over time. In these simulations, each user (node) has a propensity to produce content, which can be influenced by both intrinsic factors and the activity of their neighbors. By assigning users to treatment and control groups while accounting for their network connections, we can run numerous simulations to observe how treatment effects propagate through the network.

These simulations reveal several key effects that impact A/B testing on social networks. One such effect is the spillover effect, where the treatment effect spreads from treated users to their neighbors in the control group. Related to this is the experiment dampening effect; due to spillover, the observed difference between the treatment and control groups may be smaller than the actual effect of the treatment: the increased activity in the control group narrows the gap we aim to measure. Another phenomenon is the intrinsic dampening effect, where the full potential of the treatment is not realized because the untreated parts of the network (the control group) have lower activity levels. This reduction diminishes the network effect for treated users, as they benefit less from interactions with their less active neighbors. The clustering effect also plays a significant role; if the treated users are more tightly clustered within the network, the network effects are amplified. A clustered treatment group can lead to a higher measured effect compared to a randomly dispersed treatment group because treated users reinforce each other's activity. Additionally, the degree distribution effect arises from variations in the number of connections (degree) each user has, influencing how they and their neighbors respond to the treatment. Users with more connections may have a disproportionate impact on spreading the treatment effect through the network. Lastly, the experiment size effect shows that as the size of the treatment group increases, the absolute content production in the network rises due to the network effects. However, the relative difference between the treatment and control groups may not increase proportionally because the spillover effect also enhances the control group's activity.

These interconnected effects highlight the complexities of conducting A/B tests in social networks, where the independence assumption does not hold, and network structures significantly influence experimental outcomes. Understanding and accounting for these effects is crucial for accurate interpretation of A/B testing results in such environments.
Without accounting for these network effects, we risk mis-estimating the treatment effects—either underestimating or overestimating them depending on the network structure and the distribution of the treatment.

By leveraging Monte Carlo simulations, we can better understand and quantify these complex network effects. Simulations enable us to model user behavior by simulating how users' content production changes over time, influenced by both intrinsic factors and the activity of their neighbors. This insight helps us adjust experimental design by informing decisions on how to assign users to treatment and control groups. For example, in simulations where treated users are tightly clustered, we might observe a higher treatment effect due to the clustering effect. Recognizing this, we first need to decide which effect we want to measure (the direct effect on treated users, the spillover onto their neighbors, or the total effect of rolling the feature out to everyone), and then choose a randomization scheme, such as assigning whole clusters of the network to treatment or control, and an estimator suited to that effect (Hudgens and Halloran, 2008). For more details on experimentation on social networks, see the article \href{https://arxiv.org/abs/2312.01607}{Monte Carlo Experiments of Network Effects in Randomized Controlled Trials}.

\section{Conclusion}

Monte Carlo simulations offer a way to visualize and internalize the behavior of statistical metrics under various scenarios. They transform abstract mathematical notions into simulated experimental outcomes, allowing us to see the probabilities unfold over thousands of runs. This not only enhances our intuition but also equips us with the confidence to design better experiments and make more informed decisions based on data.

Yet, as Wigner reportedly remarked, \textit{"It's nice to know that the computer understands the problem. But I would like to understand it too."} This sentiment underscores the importance of not relying solely on computational power but also striving for a deep, conceptual grasp of the statistical principles at play. While computers can perform vast calculations and process immense datasets, it is our responsibility to interpret the results, question assumptions, and ensure that our conclusions are both mathematically valid and practically meaningful.

The unreasonable effectiveness of Monte Carlo simulations lies not just in their computational prowess but in their ability to enhance human understanding. They serve as a bridge between the complexity of statistical theory and the practical needs of experimenters seeking actionable insights. By integrating Monte Carlo simulations into our A/B testing toolkit, we enhance our ability to design robust experiments, interpret results accurately, and make decisions that are statistically sound. 

\section{Code}

The paper's accompanying code, including all figures in the paper, is available as a Python notebook on Github. The random seed has been set to a fixed value in the code, so all numerical results are reproducible by re-running the notebook. Use the below link to access the notebook:
\href{https://github.com/mtrencseni/unr-eff-monte-carlo-ab}{\texttt{{\ssmall github.com/mtrencseni/unr-eff-monte-carlo-ab}}}

\bibliographystyle{elsarticle-harv}

\end{document}